\documentstyle[prl,aps,floats,epsf,psfig]{revtex}

\newcommand{\be}{\begin{equation}}
\newcommand{\bel}[1]{\begin{equation}\label{#1}}
\newcommand{\ee}{\end{equation}}
\newcommand{\bea}{\begin{eqnarray}}
\newcommand{\ba}{\begin{array}}
\newcommand{\eea}{\end{eqnarray}}
\newcommand{\ea}{\end{array}}

\begin{document}

\twocolumn[\hsize\textwidth\columnwidth\hsize\csname@twocolumnfalse%
\endcsname

\title{Boundary-induced phase transitions in traffic flow}
%                              %
\author{V. Popkov$^{1,2}$, L. Santen$^{3}$, A. Schadschneider$^{3}$,
and G. M. Sch\"utz $^{1}$}

\address{$1$ Institut f\"ur Festk\"orperforschung, Forschungszentrum
J\"ulich, 52425 J\"ulich, Germany}
\address{$2$ Institute for Low Temperature Physics, 310164 Kharkov, Ukraine}
\address{$3$ Laboratoire de Physique Statistique, Ecole Normale
Sup{\'{e}}rieure, 24, rue Lhomond, F-75231 Paris Cedex 05, France}
\address{$4$ Institut f\"ur Theoretische Physik, Universit\"at zu K\"oln,
D-50937 K\"oln}

\date{\today}

\maketitle

\begin{abstract}Boundary-induced phase transitions are one of the 
surprising phenomena appearing in nonequilibrium systems. 
These transitions have been found in driven systems, especially
the asymmetric simple exclusion process.
However, so far no direct observations of this phenomenon in real
systems exists. 
Here we present evidence for the appearance of such a nonequilibrium
phase transition in traffic flow occurring on highways in the
vicinity of on- and off-ramps. Measurements on a German motorway close to
Cologne show a first-order nonequilibrium phase transition between 
a free-flow phase and a congested phase. It is induced by 
the interplay of density waves (caused by an on-ramp) and a shock wave 
moving on the motorway. The full phase diagram, including the effect of 
off-ramps, is explored using computer simulations and suggests means
to optimize the capacity of a traffic network.
\end{abstract}
]
%%%%%%%%%%%%%%%%%%%%%%%%%%%%%%%%%%%%%%%%%%%%%%%%%%%%%%%%%%%%%%%%%%%%%%%%

One-dimensional physical systems with short-ranged interactions in thermal
equilibrium do not exhibit phase transitions. This is no longer true if the
action of external forces sets up a steady mass transport and drives the 
system out of equilibrium. Then boundary conditions, usually insignificant 
for an equilibrium system, can induce nonequilibrium phase transitions. 
Moreover, such phase transitions may occur in a rather wide class of 
driven complex systems, including biological and sociological mechanisms 
involving many interacting agents. 
In spite of the importance of this phenomenon and a number of theoretical 
studies \cite{Krug91,Gunter93,Oerd98,Kolo98}, it has never been observed 
directly. So far only indirect experimental evidence for a 
boundary-induced phase transition exists in older studies of the 
kinetics of biopolymerization on nucleic acid templates
\cite{MacD69,Schu97}.

In the present work we report the first direct observation of a
boundary-induced phase transition in traffic flow.
Analysis of traffic data sets
taken on a motorway near Cologne exhibits  transitions from free-flow 
traffic to congested traffic, caused by a boundary effect, viz. the
presence of an on-ramp. These transitions are characterized by a
discontinuous reduction of the average speed \cite{Neub99}.

Vehicular traffic on a motorway is controlled by a mixture of bulk and
boundary effects caused by on- and off-ramps, varying number of lanes, speed
limits, leaving aside temporary effects of changing weather
conditions and other non-permanent factors.
The fundamental characteristic of the bulk motion is the stationary
flow-density diagram, i.e.\ the fundamental diagram,
which incorporates the collective
effects of individual drivers behavior such as trying to
maintain an optimal high speed while keeping safety precautions.
The qualitative shape
of the flow-density diagram $j(\rho)$ is largely independent of the precise
details of the road and hence amenable to numerical analysis using either
stochastic lattice gas models or partial differential equations
\cite{CSS99,Hell}.

\begin{figure}[h]
\centerline{\psfig{figure=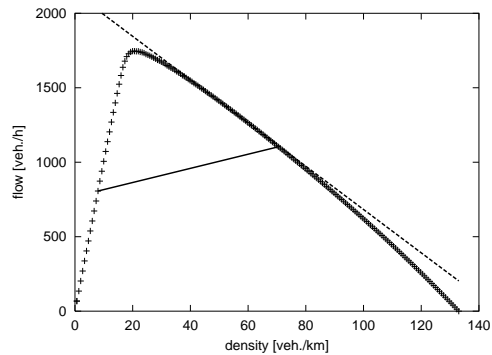,height=5cm}}
\caption{\protect{Fundamental diagram (flow-density-relation) as modeled by
the NaSch model  with  $v_{max}=4, \ p=0.25$,
time step =$1.0$ sec, lattice spacing =$7.5$ m. The system has 3200 sites,
and the flux is averaged  over  $10^6$ lattice updates. The
broken and full lines indicate the slope which defines the 
collective velocity of spontaneous local traffic jams and the shock
velocity, respectively.}}
\label{fig_fund}
\end{figure}

A by now well-established lattice gas model for traffic
flow, the cellular automaton model by Nagel and Schreckenberg \cite{NS},
reproduces empirical traffic data \cite{Hall86} rather well. 
Fig.~\ref{fig_fund} shows simulation data for the fundamental 
diagram obtained from the Nagel-Schreckenberg (NaSch) model.
This has to be compared to
measurements of the flow $j$ taken with the help of detectors on the
motorway A1 near Cologne which show a maximum of about 2000 vehicles/hour
at a density of about $\rho^\ast= 20$ vehicles per lane and km \cite{Neub99}.
At densities below $\rho^\ast$ one observes free flow, while
for larger densities one observes congested traffic.

In addition to the density dependence of the flow
two important characteristics are derived directly from the fundamental
diagram:
the shock velocity of a `domain wall' \cite{dowalldef}
between two stationary regions of densities $\rho^-,\rho^+$
\be
v_{shock} = \frac{j(\rho^+) - j(\rho^-)}{\rho^+ - \rho^-},
\label{v_shock}
\ee
obtained from mass conservation, and the collective velocity
\bel{v_c}
v_c = \frac{\partial  j(\rho) }{\partial \rho}
\ee
which is the velocity of the center of mass of a local perturbation
in a homogeneous, stationary background of density $\rho$.
Both velocities are readily observed in real traffic.
The collective velocity $v_c$ describes the upstream movement of a local,
compact jam. In the density range $25 \dots 130$ cars/km
$v_c$ ranges from approximately $-10$ km/hour to $-20$ km/hour 
(Fig.~\ref{fig_fund}) which has to be
compared with the empirically observed value $v\approx -15$ km/hour
\cite{NS,KR_Rapid}.
The shock velocity is the velocity of the
upstream front of an extended, stable traffic jam.
The formation of a stable shock is usually a boundary-driven
process, caused by a `bottleneck' on a road. Bottlenecks on a highway
arise from a reduction in the number of lanes and from on-ramps where
additional cars enter a road \cite{Daganzo,schorsch}.

\begin{figure}[h]
\centerline{\psfig{figure=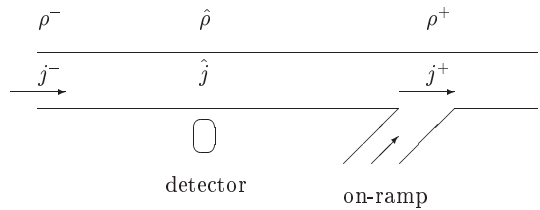,height=3.4cm}}
\caption{\protect{Schematic road design of a highway with an on-ramp
where cars enter the road. The arrows indicate the direction of the flow.
The detector measures the local bulk density $\hat{\rho}$ and bulk flow
$\hat{\jmath}$.}}
\label{fig_ramp}
\end{figure}

The experimental data considered here (see
Fig.~\ref{fig_ramp} for the relevant part of the highway) show
boundary effects caused by the presence of an on-ramp.
Far upstream from the on-ramp, free flow of 
vehicles with density $\rho^{-}$ and flow $j^-\equiv j(\rho^{-})$
is maintained. Just before the on-ramp the vehicle density is $\rho^{+}$ 
with corresponding flow $j^+\equiv j(\rho^{+})$. Note that no experimental
data are available for $\rho^-$, $j^-$ and $\rho^+$, $j^+$ as well as 
the activity of the ramp. The only data come from a detector located
upstream from the on-ramp \cite{detector_distance} which measures a 
traffic density $\hat{\rho}$ and the corresponding flow $\hat{\jmath}$.

Next the effects of the on-ramp are considered.
Cars entering the motorway cause the mainstream of vehicles
to slow down locally. Therefore, the vehicle density
just before the on-ramp increases to $\rho^{+}>\rho^{-} $.
Then a shock, formed at the on-ramp, will propagate
with {\em mean} velocity $v_{shock}$ (see (\ref{v_shock})).
Depending on the sign of $v_{shock}$, two scenarios are possible:\\
1) $v_{shock}>0$ (i.e. $j^{+}> j^{-}$): In this case
the shock propagates (on average) downstream towards the on-ramp. 
Only by fluctuations a brief upstream motion is possible. 
Therefore the detector will measure a traffic density 
$\hat{\rho}=\rho^{-}$ and flow $\hat{\jmath}= j^{-}$.\\
2) $v_{shock}<0$ (i.e. $j^{+}< j^{-}$): 
The shock wave starts propagating with the mean velocity
$v_{shock}$ upstream, thus expanding the congested traffic region with
density $\rho^{+}$. The detector will now measure $\hat{\rho}=\rho^{+}$ 
and flow $\hat{\jmath}= j^{+}$.

Let us now discuss the transition between these two scenarios.
Suppose one starts with a situation where $j^{+}> j^{-}$
is realized. If now the far-upstream-density $\rho^{-}$ increases 
it will reach a critical point $\rho_{crit} < \rho^\ast$ above which
$j^{-}> j^{+}$,
i.e., the free flow upstream $j^{-}$ prevails over the flow $j^{+}$ which 
the `bottleneck', i.e. the on-ramp, is able to support.
At this point shock wave velocity $v_{shock}$ will change sign
(see (\ref{v_shock})) and the shock starts traveling upstream.
As a result, the stationary bulk density $\hat{\rho}$ measured by
the detector upstream from the on-ramp will change discontinuously 
from the critical value $\rho_{crit}$ to $\rho^{+}$. This marks a
nonequilibrium phase transition of first order with respect
to order parameter $\hat{\rho}$. The discontinuous change 
of $\hat{\rho}$ leads also an abrupt reduction of the local velocity.
Notice that the flow $\hat{\jmath}=j^{+}$
through the on-ramp (then also measured by the detector) will stay
{\em independent} of the free flow upstream from the congested 
region $j^{-}$ as long as the condition $j^- > j^+$ holds. 

\begin{figure}[h]
\centerline{\psfig{figure=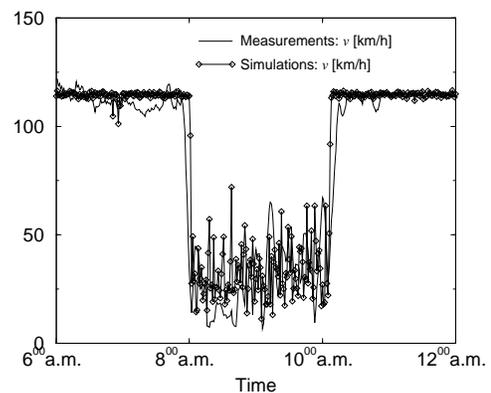,height=5.3cm}}
\caption{\protect{Time series of the velocity. Each data point represents 
an one-minute average of the speed. Shown are empirical data 
(from [7]) in comparison with the results of computer simulations
of a simplified model (see text).}}
\label{fig_time}
\end{figure}

Empirically this phenomenon can be seen in the traffic data taken from 
measurements at the detector D1 on the motorway A1 close to Cologne 
\cite{Neub99}. Fig.~\ref{fig_time} shows a typical time series of the 
one-minute velocity averages. One can clearly see the sharp drop of the 
velocity at about 8 a.m.

\begin{figure}[h]
\centerline{\psfig{figure=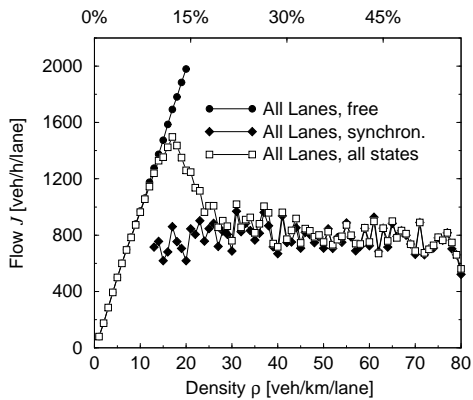,height=5.5cm}}
\caption{\protect{Measurements of the flow versus local density before an
on-ramp on the motorway A1 close to Cologne. The detector is located at a
distance 1 km upstream from on-ramp. Data are collected over a period of 12
days.}}
\label{fig_measur}
\end{figure}

Also the measurements of the flow versus local density,
i.e.\ the fundamental diagram (Fig.~\ref{fig_measur}), support our
interpretation. Two branches can be distinguished. 
The increasing part corresponds to an almost linear rise of the flow 
with density in the free-flow regime \cite{KR_Rapid}. 
In accordance with our considerations this part of the flow diagram is 
not affected by the presence of the on-ramp at all and one 
measures $\hat{\jmath} = j^{-}$ which is the actual upstream flow. 
The second branch are measurements taken during congested traffic hours, 
the transition period being omitted for better statistics. The transition 
from free flow to congested traffic is characterized by a discontinuous 
reduction of the local velocity. However, as predicted above
the flow does not change significantly in the congested regime.
In contrast, in local measurements large density fluctuations can
be observed. Therefore in this regime the density does not take the 
constant value $\rho^{+}$ as suggested by the argument given above,
but varies from 20 veh/km/lane to 80 veh/km/lane (see Fig.~\ref{fig_measur}).

One should stress here that congested traffic data are usually not easy
to interpret, because the traffic conditions (mean inflow and outflow of cars
on the on- and off-ramps, and so the bulk mean flow) are changing in time.
According to our arguments, in a congested regime the detector
measures  $j^{+}$, {\it solely} due to the on-ramp activity.
Therefore, $j^{+}(t) = j(\rho^{+}(t)) < j^{-} (t)$ must be satisfied.
During times of very dense traffic one expects always cars ready to
enter the motorway at the on-ramp, thus guaranteeing
a sufficient and approximately constant on-ramp activity.
The measured flow is constant over long periods of time which is
in agreement with the notion that the transition is due a stable
traffic jam. Spontaneously emerging and decaying jams would lead to the
observation of a non-constant flow.

The use of our approach is not limited to a qualitative explanation of 
the traffic data. Beyond that it can also be used to calculate the 
phase diagrams of systems with open boundary conditions for a large 
class of traffic models.
We modeled a section 
of a road with on-ramp on the left and off-ramp (on-ramp) on the right 
using the NaSch cellular automaton \cite{NS}.
We modify the basic model by using open boundary conditions with injection 
of cars at the left boundary (corresponding to in-flow into the road 
segment) and removal of cars at the right boundary (corresponding to 
outflow). Therefore it can also be regarded as a generalization
of the asymmetric simple exclusion process \cite{ASEP} to particles
with higher velocity.

During the simulations local measurements of the velocity have
been performed analogous to the experimental setup.
For comparison the results of the computer simulations have been
included in Fig.~\ref{fig_time}. Note that even the quantitive
agreement with the empirical data is very good. This has been achieved 
by using a finer discretization of the model, i.e.\ the 
length of the cell is considered as $l=2.5m$. The results were
obtained for $L=960$, $p=0.25$ and $v_{max}=13$. We kept the input 
probability $\alpha = 0.65$ constant. Then the free-flow part is 
obtained using $\beta = 1.$ and the congested part using 
$\beta = 0.55$. The transition was observed at ten minutes after 
we reduced the output probability. The ``detector'' was located at 
the link from site $480$ to $481$.

\begin{figure}[h]
\centerline{\psfig{figure=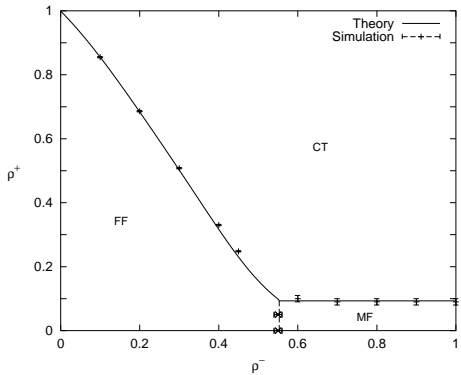,height=5.3cm}}
\caption{\protect{Phase diagram of the NaSch model with open
boundaries for $p=0.25, v_{max}=4$. Cars enter the road from a reservoir of
density $\overline{\rho^{-}}$, inducing the upstream-density $\rho^-$ discussed
in the text. At the right boundary cars leave the road into a reservoir of
density $\overline{\rho^{+}}$, leading to the on-ramp density $\rho^+$.
The solid (dashed) curve denotes the theoretical prediction for
the  first (second) order transitions lines obtained from the
numerically determined flow-density relation.  The points
represent phase transition points for the simulated system of size 3200.
The phases are: free flow (FF), congested traffic (CT), maximal flow (MF)
phase.}}
\label{fig_phase}
\end{figure}

Fig.~\ref{fig_phase} shows the full phase diagram of the NaSch model
with open boundary conditions. It describes the stationary bulk 
density $\hat{\rho}$ as a function of the far-upstream in-flow boundary 
density $\rho^-$ and the effective right boundary density $\rho^+$. 
For the case of an on-ramp (or shrinking road width etc.) at the right 
boundary corresponds to the situation discussed above. Here, the density 
is increased locally to $\rho^+ >\rho^-$. In agreement with the 
empirical observation we find a line of first order transitions 
from a free flow (FF) phase with bulk density $\hat{\rho}=\rho^-$ to a 
congested (CT) phase with $\hat{\rho}=\rho^+$. On this line 
$v_{shock}$ changes sign.

The case of an off-ramp (or expansion of road space etc.) leads to a 
local decrease $\rho^+ <\rho^-$ of the density. Here the collective 
velocity $v_c $ (\ref{v_c}) plays a prominent role. 
As long as $v_c$ is positive (i.e. in the free-flow regime
$\rho^-< \rho^\ast$, see Fig.~\ref{fig_fund}), perturbations caused by a 
small increase of the upstream boundary density $\rho^-$ gradually spread
into the bulk, rendering $\hat{\rho}=\rho^-$ (FF regime).
At $\rho^-= \rho^\ast$, $v_c$ changes sign \cite{negative_v_c}
and an overfeeding effect occurs: a perturbation from the upstream
boundary does not spread into the bulk \cite{Gunter93,Kolo98} and therefore 
further increase of the upstream boundary density does not increase the 
bulk density. The system enters the maximal flow (MF) phase with
constant  bulk density $\hat{\rho}=\rho^\ast$ and flow
$j(\rho^\ast)=j_{max}$. The transition to the MF phase is of second 
order, because $\hat{\rho}$ changes continuously across the phase
transition point.

The existence of a maximal flow phase was not emphasized in the
context of traffic flow up to now. At the same time, it is the most
desirable phase, carrying the maximal possible throughput of vehicles
$j_{max}$.  
For practical purposes our observations may directly be used in
order to operate a highway in the optimal regime. E.g.\ the flow near a
lane reduction could be increased significantly if the traffic state
at the entry would allow to attain the maximal possible flow 
of the bottleneck. This could be achieved by controlling the density 
far upstream, e.g.\ by closing temporally an on-ramp, such 
that the cars still enter the bottleneck with high velocity.

We stress that the stationary phase diagram Fig.~\ref{fig_phase} is 
generic in the sense that it is determined solely by the macroscopic
flow-density  relation.
The number of lanes of the road, the distribution of individual
optimal velocities, speed limits, and other details enter only
in so far as they determine the exact values characterizing the flow-density
relation for that particular road.
We also note that throughout the paper we assumed the external 
conditions to vary slowly, so that the system has enough time to readjust 
to its new stationary state.  Experimenting with different cellular
traffic models in a real time scale shows that the typical time to reach a
stationary state in  a road segment of about 1.2 km is
of the order of 3-5 min,
which is reasonably small. Close to phase transitions lines,
however, where the shock velocity vanishes, this time diverges
and  intrinsically non-stationary dynamic phenomena \cite{SH,RH}  take
the lead.

In conclusion, we have shown that traffic data collected on German
motorways provide evidence for a boundary-induced nonequilibrium phase
transition of first order from the free flowing to the congested
phase. The features of this phenomenon are readily understood in terms
of the flow-density diagram. The dynamical mechanism leading to this
transition is an interplay of shocks and local fluctuations caused by
an on-ramp.  Full investigation of a cellular automaton model for
traffic flow reproduces this phase transition, but also exhibits a
richer phase diagram with an interesting maximal flow phase.
These results are not only important from the point of view of
nonequilibrium physics, but also suggest new mechanisms
of traffic control.

{\bf Acknowledgments}: We thank Lutz Neubert for useful discussions and
help in producing Figs.~\ref{fig_time} and
\ref{fig_measur}. L.~S. acknowledges support from the Deutsche
Forschungsgemeinschaft under Grant No. SA864/1-1.

\bibliographystyle{unsrt}

\begin{thebibliography}{99}

\bibitem{Krug91}
Krug, J., Phys. Rev. Lett. {\bf 67}, 1882 (1991).

\bibitem{Gunter93}
 Sch\"utz, G. and Domany, E., J. Stat. Phys. {\bf 72}, 277 (1993).

\bibitem{Oerd98}
Oerding, K., and Janssen, H.K., Phys. Rev. E {\bf 58}, 1446 (1998).

\bibitem{Kolo98}
Kolomeisky, A.B., Sch\"utz, G.M.,  Kolomeisky, E.B. and  Straley, J.P.,
J.Phys. A {\bf 31}, 6911 (1998).

\bibitem{MacD69}
MacDonald, J.T. and Gibbs J.H., Biopolymers {\bf 7}, 707 (1969).

\bibitem{Schu97}
Sch\"utz, G.M., Int. J. Mod. Phys. B {\bf 11}, 197 (1997).

\bibitem{Neub99}
 Neubert, L., Santen, L.,  Schadschneider, A. and  Schreckenberg, M.,
 Phys. Rev. E {\bf 60}, 6480 (1999).

\bibitem{CSS99}
Chowdhury, D., Santen, L., and Schadschneider, A.,
Curr. Sci. {\bf 77}, 411 (1999) and Physics Reports (in press).


\bibitem{Hell}
Helbing, D., {\em Verkehrsdynamik: Neue Physikalische Modellierungskonzepte}
(in German) (Springer, Berlin, 1997).


\bibitem{NS}
Nagel, K. and  Schreckenberg, M., J. Phys. I France  {\bf 2}, 2221 (1992).

\bibitem{Hall86}
Hall, F.L., Allen, B.L. and Gunter, M.A., Transp. Res. A {\bf 20}, 197 (1986).

\bibitem{dowalldef}
In nonequilibrium systems a domain wall is an object connecting
two possible stationary states.

\bibitem{KR_Rapid} Kerner, B.S. and   Rehborn, H., Phys. Rev. E {\bf 53},
 R4275 (1997). For the situation studied in this paper the front
velocity of the jam can be identified with its center-of-mass velocity.


\bibitem{Daganzo} Daganzo, C.F., Cassidy, M.J. and Bertini R.L.,
Transp. Res. A {\bf 33}, 365 (1999).

\bibitem{schorsch} G.\ Diedrich, L.\ Santen, A.\ Schadschneider and
J.\ Zittartz, Int.\ J.\ Mod.\ Phys.\ C (in press)

\bibitem{detector_distance} The distance between the detector and the on-ramp
should be large enough, so that the on-ramp fluctuations are not
 measured directly. In our case,
the detector is located approximately 1 km upstream from on-ramp.

\bibitem{ASEP} see e.g.\ G.M. Sch\"utz, {\it Exactly solvable models for
many-body systems far from equilibrium}, to appear in {\it Phase Transitions
and Critical Phenomena}, C. Domb und J. Lebowitz (eds.), 
(Academic Press, London, 2000); T.M. Liggett, {\it Stochastic Interacting
Systems: Contact, Voter and Exclusion Processes} (Springer, Berlin, 1999);
and the contributions by B.\ Derrida, M.R\ Evans and S.A.~Janowsky,
J.L.~Lebowitz in V. Privman (ed.), {\it Nonequilibrium Statistical 
Mechanics in One Dimension}, (Cambridge University Press, Cambridge, 1997).


\bibitem{negative_v_c} In this case the upstream
entrance to the road itself acts as a `dynamical' bottleneck with maximal
capacity $j_{max}$.

\bibitem{SH} Kerner, B.S. and   Rehborn, H., Phys. Rev. Lett. {\bf 79},
 4030 (1997).

\bibitem{RH} Lee, H.Y., Lee, H.-W. and  Kim, D.,
Phys. Rev. Lett. {\bf 81}, 1130 (1998).

\end{thebibliography}

\end{document}